\documentclass[conference]{IEEEtran}
\IEEEoverridecommandlockouts
\usepackage{cite}
\usepackage{amsmath,amssymb,amsfonts}
\usepackage{algorithmic}
\usepackage{graphicx}
\usepackage{textcomp}
\usepackage{xcolor}
\usepackage{booktabs}
\usepackage{cleveref}
\def\BibTeX{{\rm B\kern-.05em{\sc i\kern-.025em b}\kern-.08em
    T\kern-.1667em\lower.7ex\hbox{E}\kern-.125emX}}
\begin{document}

\title{Interference Mitigation Recommender System using U-Net Autoencoders\\
\thanks{This paper is based upon work supported in part by
the National Science Foundation under Grant ECCS-2029948.}
}

\author{\IEEEauthorblockN{Hiten Prakash Kothari}
\IEEEauthorblockA{\textit{Wireless@VT} \\
\textit{Virginia Tech}\\
Blacksburg, VA \\
hitenkothari@vt.edu}
\and
\IEEEauthorblockN{R. Michael Buehrer}
\IEEEauthorblockA{\textit{Wireless@VT} \\
\textit{Virginia Tech}\\
Blacksburg, VA \\
rbuehrer@vt.edu}
}

\maketitle

\begin{abstract}
Building on the previous work on interference mitigation, this paper introduces a modular recommender system that automatically selects the most effective interference mitigation strategy based on the interference characteristics present in the received signal. The system integrates three key stages: an SPS classifier module, a SIR predictor, and a bank of specialized U-Net autoencoders designed for different interference conditions. The classification block identifies the parameters required for cancellation. The recommender then directs the signal to the appropriate mitigation model, optionally incorporating SIR-based decisions for scenarios where successive interference cancellation may be advantageous. Experiments conducted across diverse SIR levels and modulation environments show that the recommender strategy improves robustness and reduces BER compared to using any single mitigation method alone. The results demonstrate the potential of adaptive, model-selective architectures to enhance interference resilience in dynamic communication environments.
\end{abstract}


\section{Introduction}

Modern wireless communication systems are vital to everyday connectivity, from mobile phones and Wi-Fi networks to satellite and military communications. As the demand for faster and more reliable data transmission grows, so does the complexity of the radio frequency (RF) environment. In many practical scenarios, multiple signals are transmitted over shared frequency bands, which leads to interference, a major challenge that directly impacts the performance of communication systems.

One of the most commonly used single carrier digital modulation schemes in wireless systems is Quadrature Phase Shift Keying (QPSK). QPSK has been used as a representative example for majority of this research for convenience. QPSK is favored for its balance between spectral efficiency and robustness to noise. However, when multiple QPSK signals are transmitted simultaneously, such as in adjacent channels or overlapping time slots, inter-signal interference becomes a significant problem. This interference can distort the signal of interest, causing errors during demodulation and degrading overall system reliability. This poses challenges for traditional communication systems, where spectral congestion reduces link reliability.

Traditional signal processing methods like filtering, successive interference cancellation (SIC), and adaptive equalization have been widely used to mitigate such interference. While these methods work well under ideal conditions, they often struggle when the interference is nonlinear or time-varying. This becomes even more difficult when the interference and the signal of interest use the same modulation scheme, making them hard to separate.

In recent years, deep learning has shown great potential in learning complex patterns and recovering meaningful signals from noisy or distorted inputs. Unlike conventional algorithms that rely on predefined rules, deep learning models can be trained to automatically discover the underlying structure of the data. This makes them promising tools for challenging signal processing tasks, including interference mitigation.

Survey work by Oyedare et al. provides a broad review of deep learning methods for interference suppression, identifying challenges such as generalization and real-time deployment \cite{oyedare_interference_2022}. Lancho et al. introduce the RF Challenge dataset and benchmark, showing that UNet and WaveNet architectures achieve orders-of-magnitude BER improvements over classical filtering and linear estimators, and highlighting the potential of data-driven interference rejection approaches \cite{lancho_rf_2024}. For radio communication denoising, Almazrouei et al. use a convolutional denoising autoencoder for spectrogram-based de-noising across IEEE 802.11 protocols \cite{almazrouei_deep_2019}, while Çağkan et al. adapt the DEMUCS architecture with a U-Net and LSTM bottleneck for RF denoising, achieving strong BER and MSE improvements \cite{yapar_demucs_2024}.

Successive interference cancellation (SIC) has also been reimagined with deep learning. Luong et al. propose SICNet, a DNN-based alternative to classical SIC, robust to CSI uncertainty and user variations \cite{van_luong_deep_2022}, while Goh et al. extend this to ISICNet for iterative cancellation in NOMA, showing resilience against imperfect CSI \cite{yun_goh_iterative_2024}.

This paper focuses only on the case where both the Signal of Interest and the Interference are QPSK signals. It is assumed that there is no frequency offset and timing offset in these datasets. The analysis conducted in the previous work on interference mitigation revealed that different interference cancellation methods exhibit varying effectiveness across different Signal-to-Interference Ratio (SIR) conditions. Specifically, the convolutional autoencoder models demonstrated superior performance in certain SIR ranges, whereas the traditional Successive Interference Cancellation (SIC) method proved advantageous in other scenarios (mostly in lower SIR ranges). Consequently, there is a clear need for a systematic approach capable of dynamically selecting the optimal interference mitigation method based on real-time interference conditions. To address this, an Interference Cancellation Recommender System was proposed and developed, integrating multiple classifier stages to recommend the most effective mitigation technique, thus optimizing overall system performance. This system is termed as \textbf{Full Successive Interference Cancellation using U-Net} or \textbf{Full SICU-Net}.

\section{System Architecture}

\begin{figure}[!htb]
    \centering
    \includegraphics[width=1\linewidth]{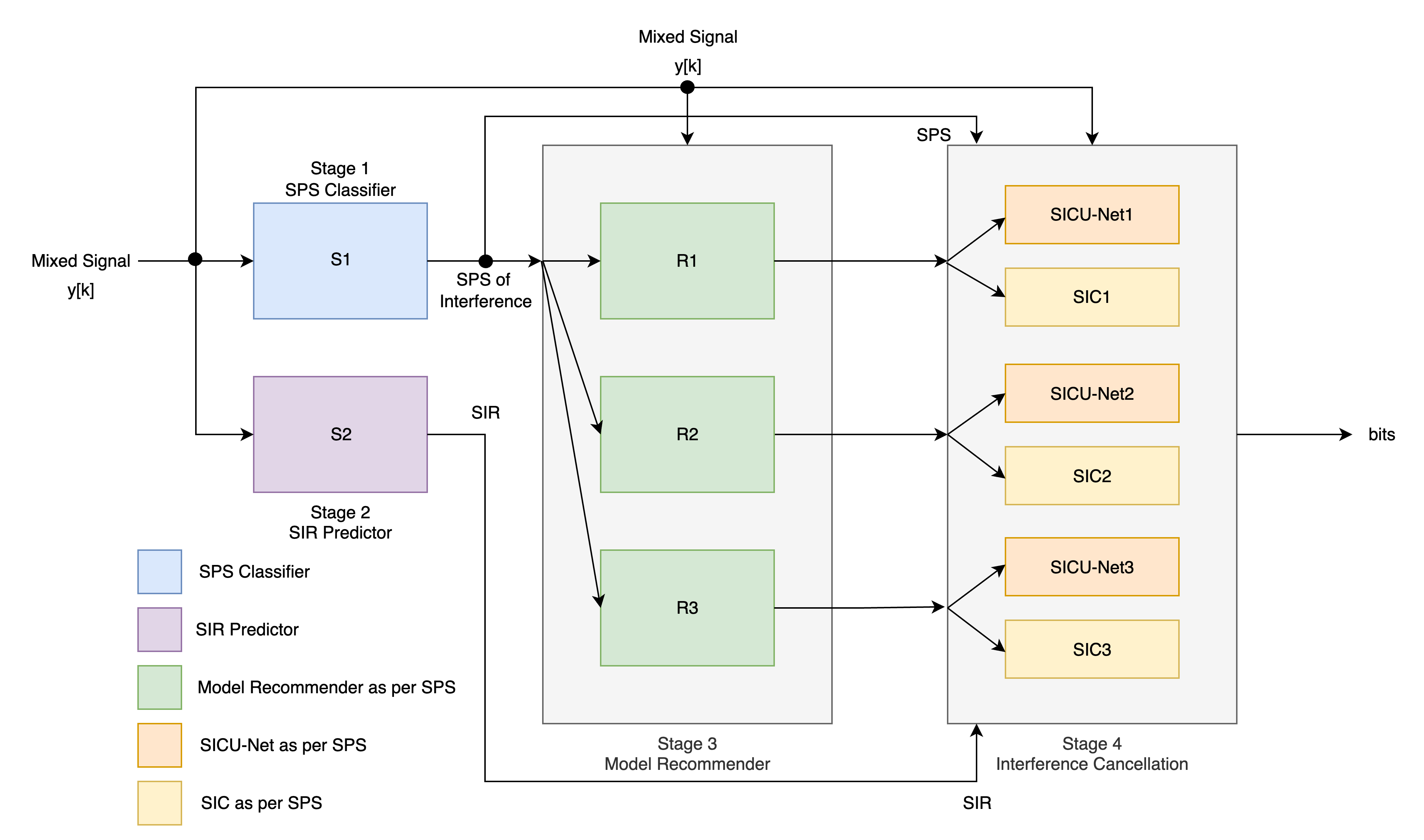}
    \caption{Recommender System Architecture}
    \label{fig:recommender}
\end{figure}

The recommender system (\Cref{fig:recommender}) comprises a four-stage pipeline designed to intelligently select and execute the most suitable interference mitigation strategy based on input signal characteristics and estimated parameters:

\begin{enumerate}
\item \textbf{Stage 1: Samples per Symbol (SPS) Classifier}: This initial stage employs a CNN-based classifier trained explicitly to predict the SPS of the interfering signal from the received mixed waveform. Accurate SPS classification enables effective subsequent parameter estimation and mitigation method selection.

\item \textbf{Stage 2: Signal-to-Interference Ratio (SIR) Predictor}: Utilizing another CNN-based classifier, this stage accurately predicts the discrete SIR value of the received signal within the predefined range of -10 dB to +10 dB. Precise SIR estimation is critical for choosing the optimal interference mitigation approach and for the scaling of the interference for cancellation.

\item \textbf{Stage 3: Model Recommender}: Given the SPS information from Stage 1 and the mixed signal waveform, this stage implements a classification model designed to recommend either the convolutional autoencoder-based approach or the traditional SIC method for interference cancellation. The recommendation model is trained to identify the optimal mitigation technique based on historical performance across different SPS and SIR conditions.

\item \textbf{Stage 4: Interference Cancellation and Bit Recovery}: Depending on the recommendation from Stage 3, this final stage employs either the U-Net based autoencoder \cite{ronneberger_u-net_2015} or the SIC approach. Utilizing the estimated SIR information from Stage 2 and SPS from Stage 1, the selected method processes the mixed signal waveform to recover the transmitted bits, yielding the final reconstructed signal.
\end{enumerate}

\section{CNN Recommender}
\subsection{Architecture}
\begin{figure}[!htb]
    \centering
    \includegraphics[width=1\linewidth]{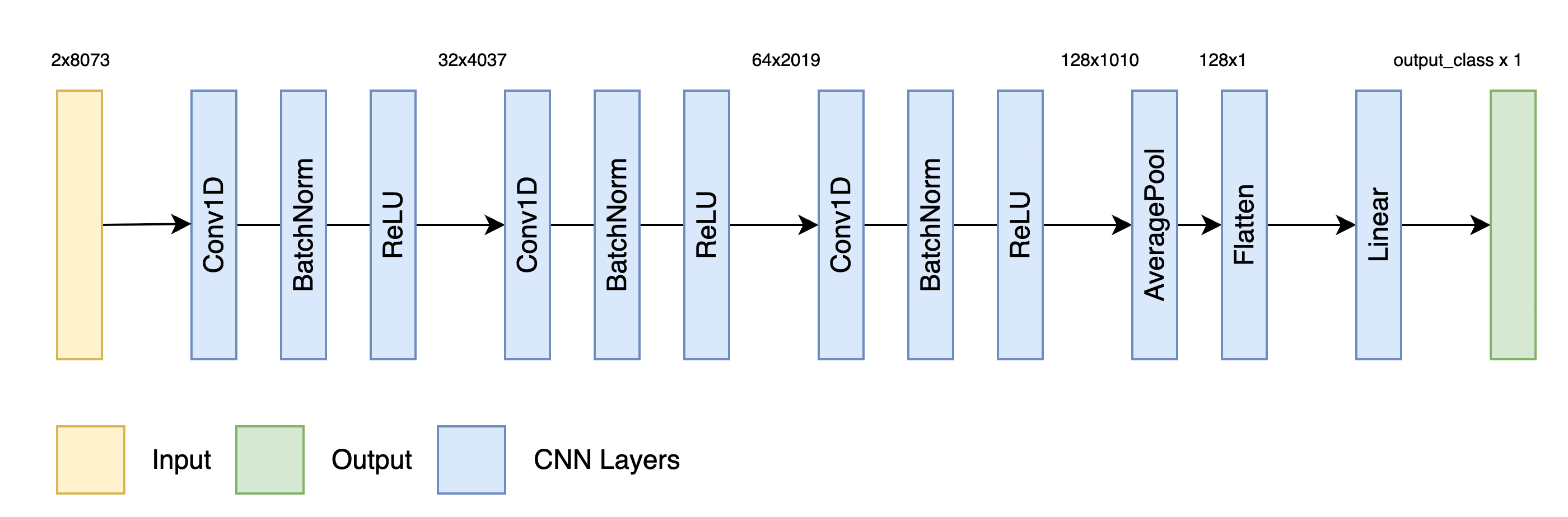}
    \caption{CNN Classifier Architecture}
    \label{fig:cnn_recommender}
\end{figure}

The CNN recommender is the classifier implemented as a convolutional neural network designed to process the in-phase (I) and quadrature (Q) components of the received baseband signal as a two-channel time series. This classifier structure is used in all three recommender stage of SPS classifier, SIR predictor, and Model recommender. The fully connected layer in these classifiers then projects this representation to the output classes as defined by the particular stage which would be 3 for the SPS Classifier (32, 16, 4), 21 values for the SIR Predictor (-10 to 10 dB), and 2 for the two models of the model classifier (SIC or SICU-Net). \Cref{fig:cnn_recommender} shows the architecture implemented for these CNN classifiers.

\subsection{Training}  
The classifier is trained in a supervised manner using mixed-signal datasets labeled with the ground-truth as per requirement. The inputs are two-channel sequences of length 8073 samples, and the outputs are categorical labels identifying the SPS/SIR/Model class. Cross-entropy loss is used as the objective function, and optimization is carried out using the Adam optimizer \cite{adam2014method}. Minibatch training is employed to accelerate convergence, and batch normalization layers stabilize training by normalizing intermediate activations. The network outputs posterior probabilities across the classes, with the final prediction taken as the class with maximum probability. Performance is evaluated using accuracy and confusion matrices across a range of SIR/SPS values to verify robustness. For training the model recommender, a ground truth table was created using the bit error rate obtained from both the methods from the work done in QPSK + QPSK case and choosing the model which provided the lower BER between the two.

\section{Results and Evaluation}

To evaluate the performance of the interference mitigation recommender system, two experimental conditions were considered. In the first case, the testing datasets were generated using integer SIR values in the range $[-10, 10]$ dB. In the second case, fractional SIR offsets were introduced, producing non-integer SIR values to test the generalization capability of the classifiers and the overall system. In each of this scenarios, 100 examples were generated for every SIR value within the range. For both cases, performance was analyzed at the level of the SPS classifier, the SIR classifier, and the final bit error rate (BER) achieved by the integrated SIC–U-Net system.

\subsection{Integer SIR}  
\begin{figure}[!htb]
    \centering
    \includegraphics[scale=0.4]{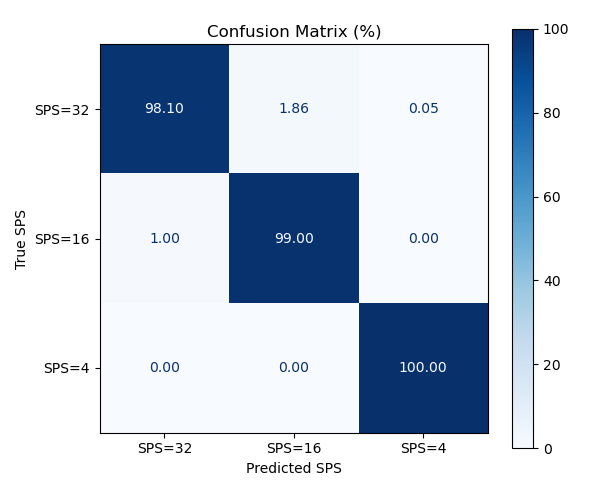}
    \caption{Stage 1: Confusion Matrix for Integer SIR} 
    \label{fig:sps_classifier11}
\end{figure}

\begin{figure}[!htb]
    \centering
    \includegraphics[scale=0.4]{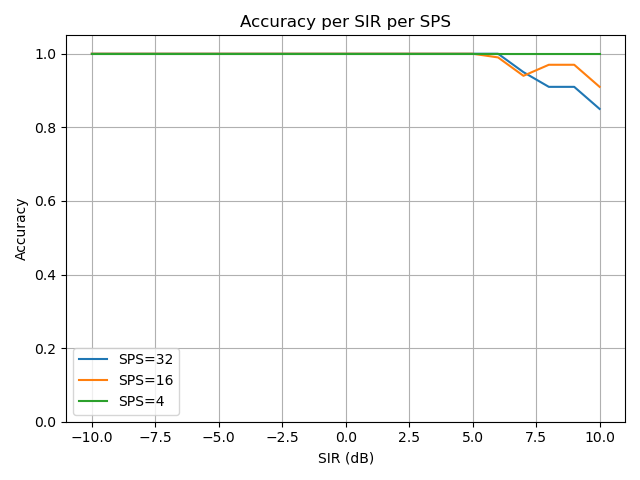}
    \caption{Stage 1: SPS Classifier for Integer SIR} 
    \label{fig:sps_classifier12}
\end{figure}

\begin{figure}[!htb]
    \centering
    \includegraphics[scale=0.25]{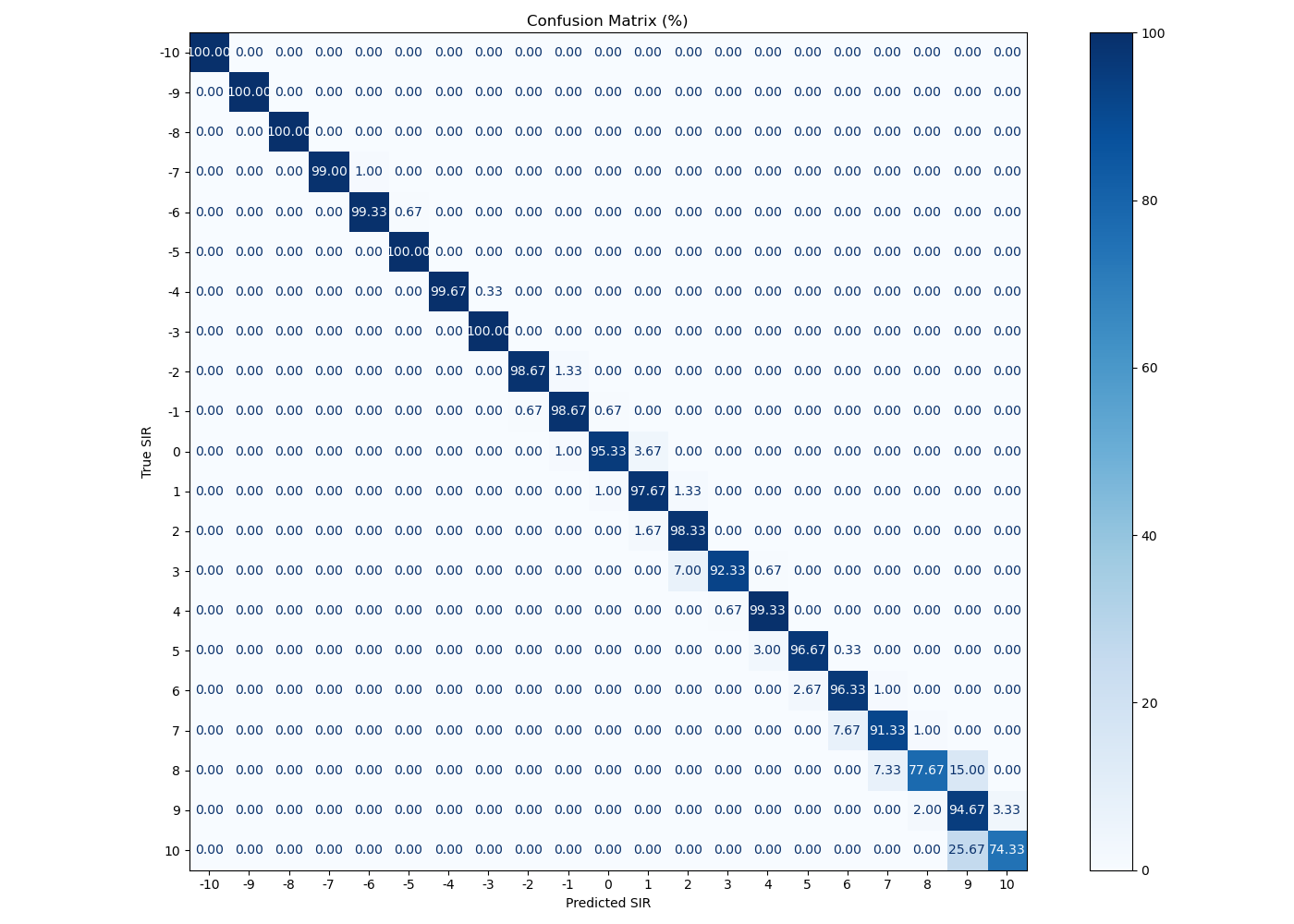}
    \caption{Stage 2: SIR Predictor for Integer SIR} 
    \label{fig:sir_predictor1}
\end{figure}

\begin{figure}[!htb]
    \centering
    \includegraphics[scale=0.5]{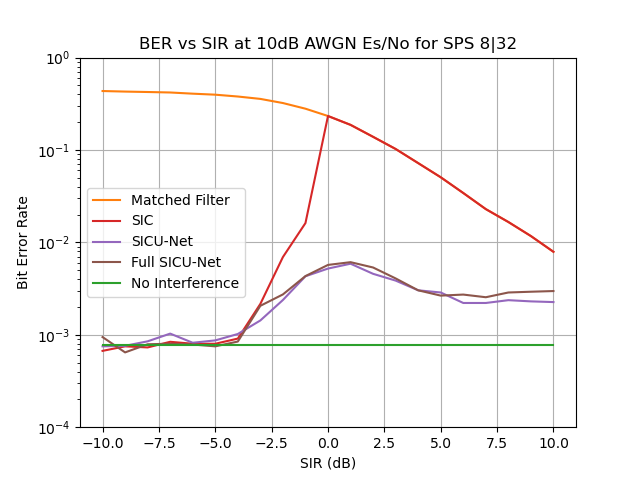}
    \caption{SPS 32 for Integer SIR} 
    \label{fig:full_ber_qpsk11}
\end{figure}

\begin{figure}[!htb]
    \centering
    \includegraphics[scale=0.5]{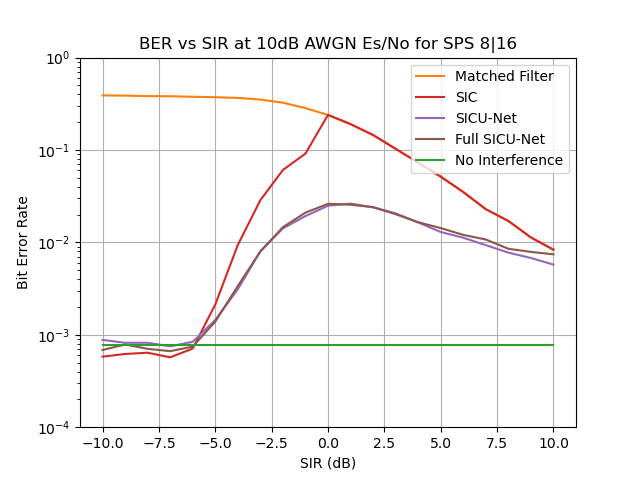}
    \caption{SPS 16 for Integer SIR} 
    \label{fig:full_ber_qpsk12}
\end{figure}

\begin{figure}[!htb]
    \centering
    \includegraphics[scale=0.5]{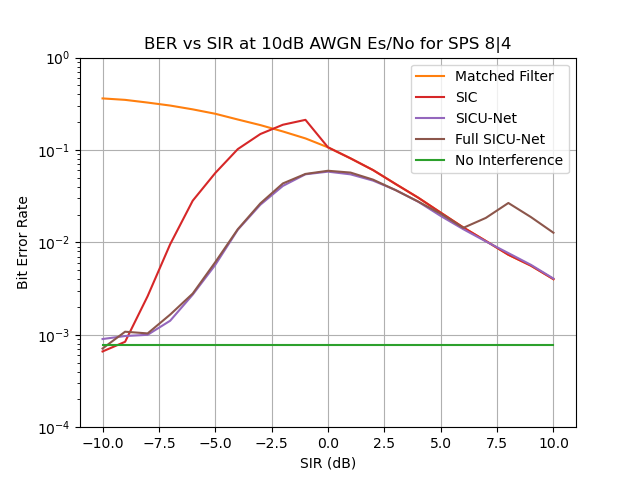}
    \caption{SPS 4 for Integer SIR}
    \label{fig:full_ber_qpsk13}
\end{figure}

When datasets were generated with integer-valued SIR levels, the SPS classifier achieved high accuracy across all three classes ($\{32, 16, 4\}$). The overall testing accuracy for the SPS classifier was 99.11\%. The resulting confusion matrices, shown in \Cref{fig:sps_classifier11,fig:sps_classifier12} indicate clear separability between the different SPS values, even under low-SIR conditions. The SIR classifier also produced an overall accuracy of 95.56\%, shown in \Cref{fig:sir_predictor1}, with most predictions lying on or near the correct SIR class. These accurate intermediate predictions allows the recommender system to select the appropriate mitigation strategy reliably. The final BER curves of the Full SIC–U-Net system shown in \Cref{fig:full_ber_qpsk11,fig:full_ber_qpsk12,fig:full_ber_qpsk13} demonstrate improvements over using either method alone, as it follows the best of both methods as well as automatically choosing the models as per required SPS, with performance gains most pronounced in the intermediate SIR regime where the recommender frequently switches between SIC and U-Net based on estimated conditions.






\subsection{Fractional SIR}  
\begin{figure}[!htb]
    \centering
    \includegraphics[scale=0.25]{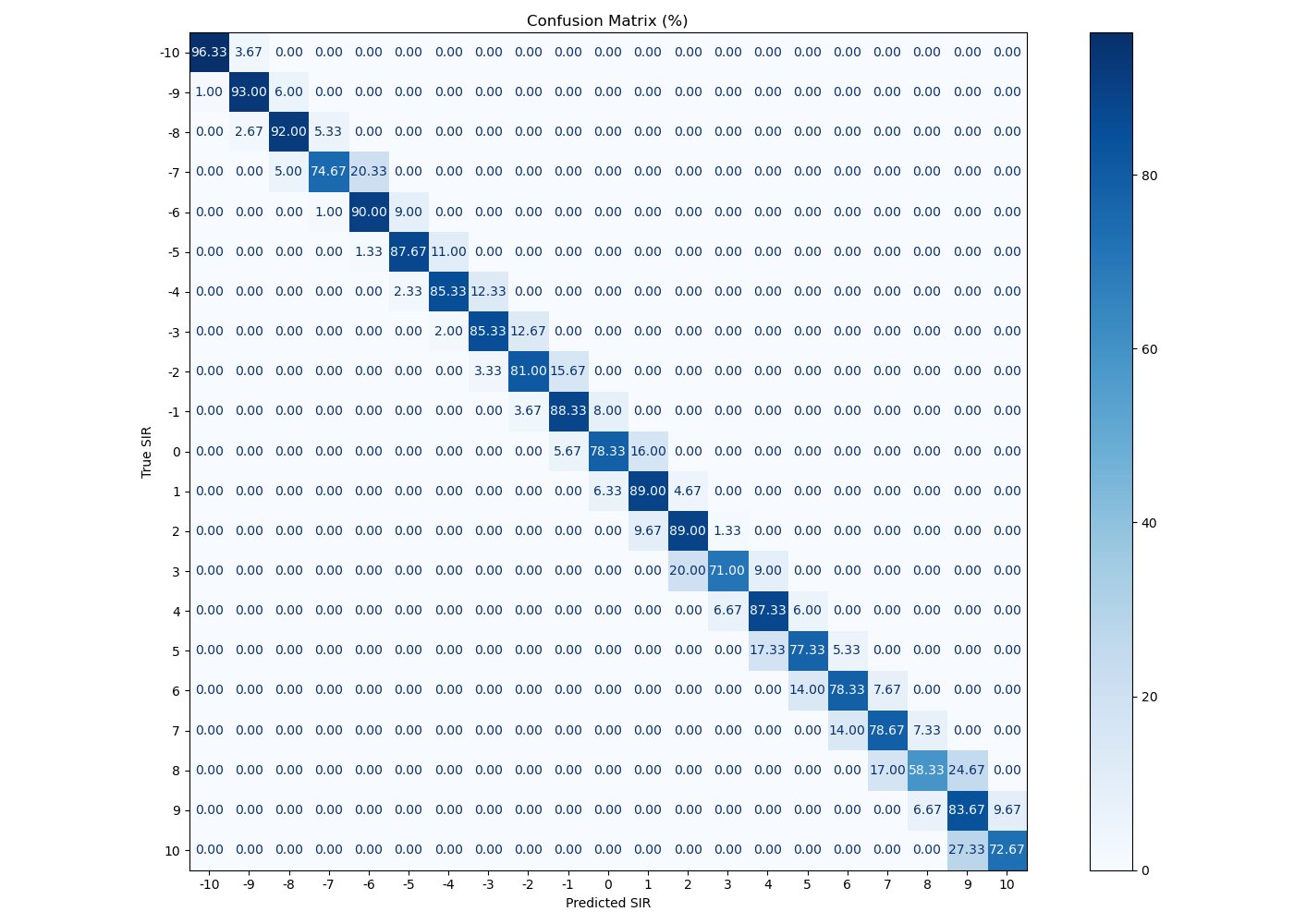}
    \caption{Stage 2: SIR Predictor for Fractional SIR} 
    \label{fig:sir_predictor2}
\end{figure}

\begin{figure}[!htb]
    \centering
    \includegraphics[scale=0.5]{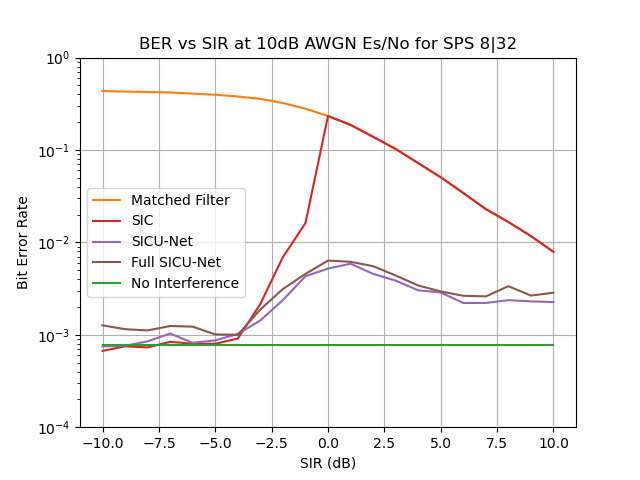}
    \caption{SPS 32 for Fractional SIR} 
    \label{fig:full_ber_qpsk21}
\end{figure}

\begin{figure}[!htb]
    \centering
    \includegraphics[scale=0.5]{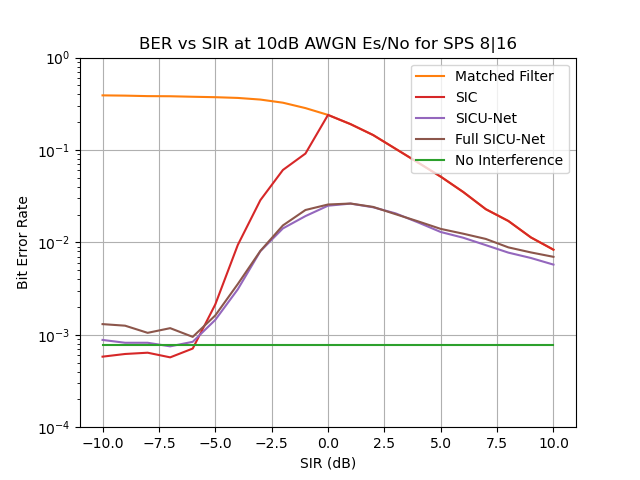}
    \caption{SPS 16 for Fractional SIR} 
    \label{fig:full_ber_qpsk22}
\end{figure}

\begin{figure}[!htb]
    \centering
    \includegraphics[scale=0.5]{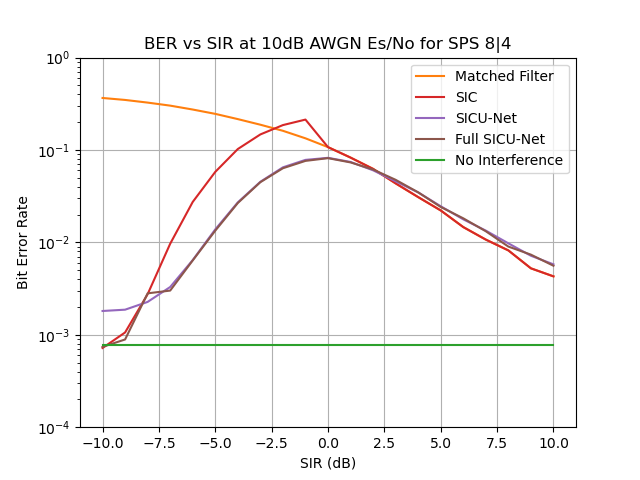}
    \caption{SPS 4 for Fractional SIR}
    \label{fig:full_ber_qpsk23}
\end{figure}

To test the system’s robustness in more realistic scenario, datasets were generated with SIR values including fractional offsets. The offset to add was generated from a uniform distribution ranging from $-0.5$ dB to $0.5$ dB. This setting introduces additional complexity since the classifiers were originally trained on integer SIR values. The SPS classifier maintained high accuracy of 98.90\%, showing that symbol-rate discrimination is largely unaffected by fractional SIR levels. The SIR classifier, however, showed increased confusion between adjacent integer bins, as expected with accuracy dipping to 82.35\%. Nonetheless, the recommender system proved resilient: by mapping fractional predictions to the nearest SIR class, it continued to select the appropriate mitigation strategy in most cases. As can be observed from \Cref{fig:sir_predictor2} the predictor estimates a SIR bin closest to the true bin, which results in a very similar output with a slight degradation because of incorrect scaling during cancellation. The final BER curves again confirmed that the SIC–U-Net system outperforms either baseline across all tested SPS values, with only minor degradation compared to the integer-SIR case owing to incorrect cancellation because of adjacent class SIR. The SIC and SICU-Net model which are used for the BER curves in \Cref{fig:full_ber_qpsk21,fig:full_ber_qpsk22,fig:full_ber_qpsk23} are for the integer SIR case and therefore they perform marginally better than the Full SICU-Net system.


\begin{figure}[!htb]
    \centering
    \includegraphics[scale=0.4]{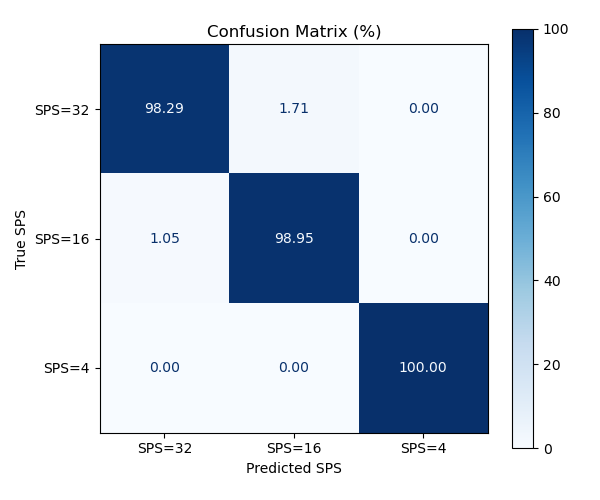}
    \caption{Stage 1: Confusion Matrix for Fractional SIR} 
    \label{fig:sps_classifier21}
\end{figure}

\begin{figure}[!htb]
    \centering
    \includegraphics[scale=0.4]{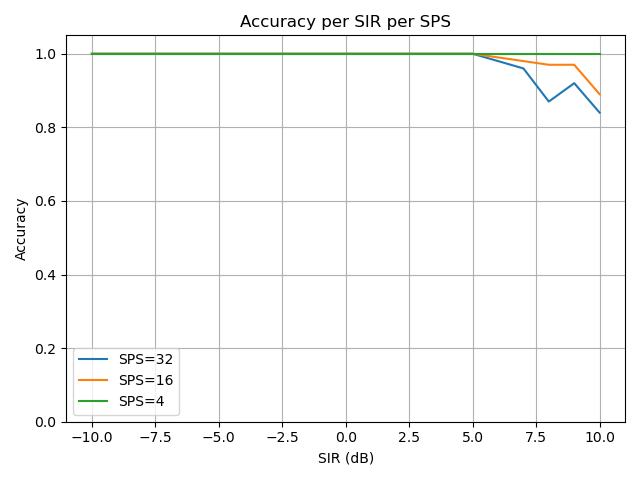}
    \caption{Stage 1: SPS Classifier for Fractional SIR} 
    \label{fig:sps_classifier22}
\end{figure}


\section{Conclusions}

This paper presented the design and evaluation of an interference mitigation recommender system that dynamically selects between a U-Net based denoising model and a successive interference cancellation (SIC) approach. The focus was on QPSK signals corrupted by QPSK interference with varying relative bandwidth, where previous results had shown that neither method consistently dominates across all interference power regimes. By integrating SPS and SIR classifiers into the decision pipeline, the system is able to adapt its mitigation strategy to the observed conditions.

The results demonstrated that SPS classification is highly reliable across all tested interference scenarios, while SIR classification provides sufficient resolution to guide the choice between SIC and SICU-Net. When tested with datasets containing both integer-valued SIR levels and fractional offsets, the system maintained robust performance. The use of confusion matrices for both SPS and SIR classification highlighted the accuracy of the intermediate stages, while final BER curves showed that the full recommender system consistently outperformed either SIC or U-Net applied independently. In particular, the adaptive switching allowed the system to achieve near-optimal performance in regimes where one method alone would have struggled.

Overall, the recommender framework illustrates the potential of combining classical signal processing with deep learning methods in an adaptive system. By leveraging the strengths of both approaches and tailoring the mitigation strategy to the channel conditions, the system achieves improved robustness and efficiency. This lays the foundation for future extensions to more complex modulation formats, multi-user interference, and real-world wireless environments.

\bibliography{IEEEabrv,./references}
\bibliographystyle{IEEEtran}

\end{document}